\documentclass[11pt]{article}
\usepackage{amsmath}
\usepackage{amsfonts}
\usepackage{amssymb}
\usepackage{graphicx}

\def\bea{\begin{eqnarray}}
\def\eea{\end{eqnarray}}

\begin{document}
\begin{center}
\LARGE { \bf Hidden Conformal Symmetry of Rotating Black Holes in Minimal Five-Dimensional Gauged Supergravity
  }
\end{center}
\begin{center}
{\bf M. R. Setare\footnote{rezakord@ipm.ir} \\  V. Kamali\footnote{vkamali1362@gmail.com}}\\
 {\ Department of Science of Bijar, University of  Kurdistan  \\
Bijar, IRAN.}
 \\
 \end{center}
\vskip 3cm

\begin{abstract}
In the present paper we show that for a low frequency limit the wave equation of massless scalar field in the background of non-extremal charged rotating black holes in five-dimensional minimal gauged and ungauged supergravity can be written as the Casimir of an $SL(2,R)$ symmetry, our result show that the entropy of the black hole is reproduced by the Cardy formula, also the absorption cross section is in consistence with the finite temperature absorption cross section for a 2D CFT.
\\

\end{abstract}

\newpage

\section{Introduction}
Recent investigation on the holographic dual descriptions for the black holes have achieved substantial success.  According to the  Kerr/CFT correspondence \cite{1}, the microscopic entropy of four-dimensional extremal Kerr black hole was calculated by studying the dual chiral conformal field theory associated with the diffeomorphisms of near horizon geometry of Kerr black hole. Susequently, this work was extended to the case of near-extreme black holes \cite{2} (see also \cite{3}). The main progress are made essentially on the extremal and near extremal limits in which the black hole near horizon geometries consist a certain AdS structure and the central charges of dual CFT can be obtained by analyzing
the asymptotic symmetry following the method in \cite{4} or by calculating the boundary stress tensor of the $2D$ effective action \cite{5}.
More recently, Castro, Maloney and Strominger \cite{6} have given evidence that the physics of Kerr black holes might be captured by a conformal field theory. The authors have discussed that the existence of conformal invariance in a near horizon geometry is not necessary  condition, instead the existence of a local conformal invariance in the solution space of the wave equation for the propagating field is sufficient to ensure a dual CFT description(see also \cite{9}). In the microscopic description, using the Cardy formula for the microscopic degeneracy, they reobtain the Bekenestein-Hawking entropy of the black hole.\\

 In this paper, we investigate the hidden conformal symmetry of non-extremal charged rotating black holes in five-dimensional minimal gauged and ungauged suppergravity  \cite{7}. These types of black holes characterized by four non-trivial parameters, namely the mass, the charge, and the two independent rotation parameters. The Bekenestein-Hawking entropy of black hole is recovered by the Cardy formula and the CFT temperatures was obtained respectively. Then, we investigate the absorption across section of the black hole in the near region and get the result that the near region of the rotating black hole in minimal five-dimensional gauged supergravity is dual to 2D CFT.\\ The paper is organized as follows. In section 2, at first we review some basic properties of charged rotating black hole in minimal gauge supergravity, then study the scattering process of a probe massless scalar field propagating in its background. In section 3, we study the hidden conformal symmetry of charged rotating  black hole in five-dimensional minimal gauged suppergravity by analysing the wave equation of scalar field. In section 4, we reproduce the Bekenstein-Hawking entropy by the Cardy formula of the dual conformal field. In section 5, we study the hidden conformal symmetry of charged rotating  black hole in five-dimensional minimal ungauged suppergravity.  In section 6, the absorption cross section is investigated, and finally, the last section is devoted to the conclusion. \\

\section{Scalar field in the background of five-dimensional  space-times }
In terms of Boyer-Lindquist type coordinate $x^{\mu}=(t,r,\theta,\phi,\psi)$, the author of \cite{7} have found that the metric and gauge potential for rotating solution in five-dimensional minimal supergravity can be expressed as

\begin{eqnarray}\label{1}
   ds^2=-\frac{\Delta_{\theta}[(1+g^2r^2)\rho^2dt+2q\nu]dt}{\Xi_a\Xi_b \rho^2}+\frac{2q\nu\omega}{\rho^2}\\
   \nonumber
   +\frac{f}{\rho^4}(\frac{\Delta_{\theta}dt}{\Xi_{a}\Xi_{b}}-\omega)^2+\frac{\rho^2dr^2}{\Delta_{r}}+\frac{\rho^2d\theta^2}{\Delta_{\theta}}\\
  \nonumber
   +\frac{r^2+a^2}{\Xi_{a}}\sin^2\theta d\phi^2+\frac{r^2+a^2}{\Xi_{a}}\cos^2\theta d\psi^2
\end{eqnarray}

\begin{eqnarray}\label{2}
  A=\frac{\sqrt{3}q}{\rho^2}(\frac{\Delta_{\theta}dt}{\Xi_a\Xi_b}-\omega)
\end{eqnarray}
where
\begin{eqnarray}\label{3}
  \nu=b\sin^2\theta d\phi+a\cos^2\theta d\psi~~~~~~~~~~~~~~~~~~~~~~~~~~~~~~~~~~\\
  \nonumber
  \omega=a\sin^2\theta\frac{d\phi}{\Xi_{a}}+b\cos^2\theta\frac{d\psi}{\Xi_b}~~~~~~~~~~~~~~~~~~~~~~~~~~~~~~~~~\\
  \nonumber
  \Delta_{\theta}=1-a^2g^2\cos^2\theta-b^2g^2\sin^2\theta~~~~~~~~~~~~~~~~~~~~~~~~~~\\
  \nonumber
  \Delta_r=\frac{(r^2+a^2)(r^2+b^2)(1+g^2r^2)+q^2+2abq}{r^2}-2m~\\
  \nonumber
  \rho^2=r^2+a^2\cos^2\theta+b^2\sin^2\theta~~~~~~~~~~~~~~~~~~~~~~~~~~~~~~~\\
  \nonumber
  \Xi_a=1-a^2g^2~~~~~~~~~~~~~~~\Xi_b=1-b^2g^2~~~~~~~~~~~~~~~~~\\
  \nonumber
  f=2m\rho^2-q^2+2abqg^2\rho^2~~~~~~~~~~~~~~~~~~~~~~~~~~~~~~~~~~
\end{eqnarray}
The  non-vanishing metric components are given by

\begin{eqnarray}\label{4}
  g_{00}=-\frac{\Delta_{\theta}(1+g^2r^2)}{\Xi_a\Xi_b}+\frac{\Delta_{\theta}^2(2m\rho^2-q^2+2abqg^2\rho^2)}{\rho^4\Xi_a^2\Xi_b^2}~~~~~~~~~~~\\
\nonumber
   g_{03}=-\frac{\Delta_{\theta}[a(2m\rho^2-q^2)+bq\rho^2(1+a^2g^2)]\sin^2\theta}{\rho^4\Xi_a^2\Xi_b}~~~~~~~~~~~~~~\\
\nonumber
   g_{04}=-\frac{\Delta_{\theta}[b(2m\rho^2-q^2)+aq\rho^2(1+b^2g^2)]\cos^2\theta}{\rho^4\Xi_b^2\Xi_a}~~~~~~~~~~~~~~~\\
\nonumber
   g_{33}=\frac{(r^2+a^2)\sin^2\theta}{\Xi_a}+\frac{a[a(2m\rho^2-q^2)+2bq\rho^2]\sin^4\theta}{\rho^4\Xi_a^2}~~~~~~~\\
\nonumber
   g_{44}=\frac{(r^2+b^2)\cos^2\theta}{\Xi_b}+\frac{b[b(2m\rho^2-q^2)+2aq\rho^2]\cos^4\theta}{\rho^4\Xi_b^2}~~~~~~~\\
\nonumber
   g_{34}=\frac{[ab(2m\rho^2-q^2)+(a^2+b^2)q\rho^2]\sin^2\theta\cos^2\theta}{\rho^4\Xi_a\Xi_b}~~~~~~~~~~~~~~~\\
\nonumber
  g_{11}=\frac{\rho^2}{\Delta}~~~~~~~~~~~~~~g_{22}=\rho^2~~~~~~~~~~~~~~~~~~~~~~~~~~~~~~~~~~~~~~~
\end{eqnarray}

 For the above metric $\sqrt{-g}=\frac{r\rho^2\sin\theta\cos\theta}{\Xi_a\Xi_b}$, the event horizons of the space-time are given by the singularities of the metric function
which are the real roots of $r^2\Delta=0$.
In the above metric $q$ and $M$ are respectively charge  and mass of black hole.
Now we consider a bulk massless scalar field $\Phi$ propagating in the background of (\ref{1}). The Klein- Gordon equation for massless scalar field is

\begin{equation}\label{5}
\Box\phi=\frac{1}{\sqrt{-g}}\partial_{\mu}(\sqrt{-g}g^{\mu\nu}\partial_{\nu})\Phi=0
\end{equation}

The wave equation can be simplified by assuming the following form of the scalar field

\begin{eqnarray}\label{6}
  \Phi(t,r,\theta,\varphi,\psi)=\exp(-i\omega t+i m_{\varphi}\varphi+i m_{\psi}\psi)S(\theta)  R(r)
\end{eqnarray}

The near region, which is the crucial region for demonstrating the origin of conformal structure, is defined by

\begin{eqnarray}\label{7}
 r\ll\frac{1}{\omega}~~~~~~~~~~~~~~~~M\ll\frac{1}{\omega}
\end{eqnarray}

 For  black holes in minimal five-dimensional supergravity  (with $g=0$), if we focus on near region, the radial equation should be rewritten in terms of the $SL(2,R)$ quadratic Casimir . The same treatment does not work in the   black holes in minimal five-dimensional gauged  supergravity,  as the function $\Delta_r$ is not  quadratic. However, the situation is  similar to the Kerr-Newman-AdS-dS  and Kerr-Bolt-AdS-dS cases. Following the treatment in \cite{chen},  we expand the function $\Delta_r$  in the near horizon region to the quadratic order of $(r^2-r_+^2)$,

\begin{eqnarray}\label{8}
r^2\Delta(r^2)\simeq K(r^2-r_{+}^2)(r^2-r_{*}^2)
  \end{eqnarray}
with
\begin{eqnarray}\label{9}
K=3g^2r_{+}^2+g^2(a^2+b^2)+1
  \end{eqnarray}
and
\begin{eqnarray}\label{10}
r_{*}^2=r_{+}^2-\frac{1}{K}(3g^2r_{+}^4+2(g^2(a^2+b^2)+1)r_{+}^2+a^2+b^2+a^2b^2g^2-2M)
  \end{eqnarray}
  In general, $r_{*}$ is not the inner horizon. Only in the case that $\Delta_r$  is quadratic, which happens in the black holes in minimal five-dimensional supergravity   case, $r_{*}$ coincides with the other horizon. In the low frequency limit and in near horizon region the radial equation for black holes in minimal five-dimensional gauged  supergravity should be simplified even more  for $m_{\psi}=0$

\begin{eqnarray}\label{11}
   [\partial_{u}(u-u_+)(u-u_{*}) \partial_{u})+\frac{u_{+}-u_{*}}{4(u-u_{+})}(\frac{\omega}{Kk_{+}}-\frac{2\Xi_a\Omega_{a}}{Kk_{+}}m_{\varphi})^2\\
   \nonumber
   -\frac{u_{+}-u_{*}}{4(u-u_{*})}(\frac{\omega}{Kk_{-}}
    -\frac{2\Xi_a\Omega_{b}}{Kk_{+}}m_{\varphi})^2]\Phi=l(l+1)\Phi
\end{eqnarray}
where

\begin{eqnarray}\label{12}
u=r^2~~~~~~~u\Delta=K(u-u_{+})(u-u_{-})
~~~~~u_{+}=r_{+}^2~~~~~~~~~~~~u_{-}=r_{-}^2
\end{eqnarray}

\begin{equation}\label{13}
\Omega_{a}=\frac{a(r_{+}^2+b^2)+bq}{(r_{+}^2+a^2)(r_{+}^2+b^2)+abq}
\end{equation}

\begin{eqnarray}\label{14}
\Omega_{b}=\frac{b(r_{+}^2+a^2)+aq}{(r_{+}^2+a^2)(r_{+}^2+b^2)+abq}
\end{eqnarray}

and

\begin{equation}\label{15}
k_{+}=\frac{r_{+}(r_{+}^2-r_{-}^2)}{(r_{+}^2+a^2)(r_{+}^2+b^2)+abq}
\end{equation}

\begin{equation}\label{16}
k_{-}=\frac{r_{+}^2(r_{+}^2-r_{-}^2)}{r_{-}((r_{+}^2+a^2)(r_{+}^2+b^2)+abq)}
\end{equation}

$\Omega_a$ and $\Omega_b$ are the angular velocities in the horizon where given by \cite{7} and respectively $k_{-}>k_{+}$. The
Hawking temperature of the black hole is determined as

\begin{equation}\label{17}
T_{H}=\frac{1}{\beta_{H}}=\frac{2\pi}{k_{+}}
\end{equation}\\

\section{ Hidden conformal symmetry of black hole in five dimensional gauged supergravity}
Following \cite 6 we now show that equation (\ref{11}) can be reproduced by the introduction of conformal coordinates.
 We will show that for massless scalar field $\Phi$, there exist a hidden $SL(2,R)_{L}\times SL(2,R)_{R}$ conformal symmetry acting on the solution space. Moreover, from the spontaneous breaking of this hidden symmetry by periodic identification of $\varphi$, we can read out the left and right temperature of dual conformal field theory.
 We introduce the conformal coordinates

\begin{equation}\label{18}
\omega^{+}=\sqrt{\frac{r^2-r_{+}^2}{r^2-r_{-}^2}}\exp(2\pi T_{R}\varphi+2n_{R}t)~~
\end{equation}

\begin{eqnarray}\label{19}
\omega^{-}=\sqrt{\frac{r^2-r_{+}^2}{r^2-r_{-}^2}}\exp(2\pi T_{L}\varphi+2n_{L}t)
\end{eqnarray}

\begin{eqnarray}\label{20}
 y=\sqrt{\frac{r_{+}^2-r_{-}^2}{r^2-r_{-}^2}}\exp(\pi( T_{R}+T_{L})\varphi+(n_{R}+n_{l})t)~~~
\end{eqnarray}

\begin{equation}\label{21}
 T_{R}=\frac{K(r_{+}-r_{-})}{2\Xi_a\pi b}, ~~~~~~T_{L}=\frac{K(r_{+}+r_{-})}{2\Xi_a\pi b }~~~
\end{equation}

\begin{equation}\label{22}
~~~~~~~n_{L}=-\frac{Kr_{+}^2}{2b}(a+b)(r_{+}+r_{-})[(r_{+}^2+a^2)(r_{+}^2+b^2)+abq]
\end{equation}

\begin{equation}\label{23}
~~~~~~~n_{R}=-\frac{Kr_{+}^2}{2b}(a-b)(r_{+}-r_{-})[(r_{+}^2+a^2)(r_{+}^2+b^2)+abq]
\end{equation}

 Following \cite{6}   we define left and right moving vectors by

\begin{eqnarray}\label{24}
H_1=i\partial_{+},~~~~
\end{eqnarray}

\begin{eqnarray}\label{25}
H_0=i(\omega^{+}\partial_{+}+\frac{1}{2}y\partial_{y}),~~~~
~
\end{eqnarray}

\begin{eqnarray}\label{26}
~H_{-1}=i((\omega^{+})^2\partial_{+}+\omega^{+}y\partial_{y}-y^2\partial_{-})~~~
\end{eqnarray}

\begin{eqnarray}\label{27}
\overline{H}_1=i\partial_{-}
\end{eqnarray}

\begin{eqnarray}\label{28}
\overline{H}_0=i(\omega^{-}\partial_{-}+\frac{1}{2}y\partial_{y})
\end{eqnarray}

\begin{eqnarray}\label{29}
\overline{H}_{-1}=i((\omega^{-})^2\partial_{-}+\omega^{-}y\partial_{y}-y^2\partial_{+})
\end{eqnarray}
Which each satisfy the $SL(2,R)$ algebra

\begin{eqnarray}\label{30}
 ~~[H_0,H_{\pm1}]=\mp i H_{\pm 1},~~~~~~~~[H_{-1},H_1]=-2iH_0
\end{eqnarray}
and

\begin{eqnarray}\label{31}
~[\overline{H}_0,\overline{H}_{\pm1}]=\mp i \overline{H}_{\pm 1},~~~~~~~~[\overline{H}_{-1},\overline{H}_1]=-2i\overline{H}_0
\end{eqnarray}
We can directly calculate all the $SL(2,R)$ generators in terms of black hole coordinates

\begin{eqnarray}\label{32}
H_1=i\exp-(2\pi T_{R}\varphi+2n_{R}t)[\sqrt{\Delta}\partial_{r}~~~~~~~~~~~~~~~~~~~~~~~~~~~~~~~~~~~~~\\
\nonumber
-\frac{(n_{L}-n_{R})(r-r_{+})+(n_{L}+n_{R})(r-r_{-})}{\pi P\sqrt{\Delta}}\partial_{\varphi}~~~~~~~~~~~~~~~\\
\nonumber
+\frac{(T_{L}-T_{R})(r-r_{+})+(T_{L}+T_{R})(r-r_{-})}{ P\sqrt{\Delta}}\partial_{t}]~~~~~~~~~~~~~~
\end{eqnarray}

\begin{eqnarray}\label{33}
H_0=i[-\frac{2n_{L}}{\pi P}\partial_{\varphi}+\frac{2T_{L}}{P}\partial_{t}]~~~~~~~~~~~~~~~~~~~~~~~~~~~~~~~~~~~~~~~
\end{eqnarray}

\begin{eqnarray}\label{34}
H_{-1}=i\exp(2\pi T_{R}\varphi+2n_{R}t)[-\sqrt{\Delta}\partial_{r}~~~~~~~~~~~~~~~~~~~~~~~~~~~~~~~~~~~~~\\
\nonumber
-\frac{(n_{L}-n_{R})(r-r_{+})+(n_{L}+n_{R})(r-r_{-})}{\pi P\sqrt{\Delta}}\partial_{\varphi}~~~~~~~~~~~~~~~\\
\nonumber
+\frac{(T_{L}-T_{R})(r-r_{+})+(T_{L}+T_{R})(r-r_{-})}{ P\sqrt{\Delta}}\partial_{t}]~~~~~~~~~~~~~~
\end{eqnarray}

and

\begin{eqnarray}\label{35}
\overline{H}_{1}=i\exp-(2\pi T_{L}\varphi+2n_{L}t)[\sqrt{\Delta}\partial_{r}~~~~~~~~~~~~~~~~~~~~~~~~~~~~~~~~~~~~~\\
\nonumber
-\frac{(n_{L}-n_{R})(r-r_{+})-(n_{L}+n_{R})(r-r_{-})}{\pi P\sqrt{\Delta}}\partial_{\varphi}~~~~~~~~~~~~~~~\\
\nonumber
+\frac{(T_{L}-T_{R})(r-r_{+})-(T_{L}+T_{R})(r-r_{-})}{ P\sqrt{\Delta}}\partial_{t}]~~~~~~~~~~~~~~
\end{eqnarray}

\begin{eqnarray}\label{36}
\overline{H}_0=i[\frac{2n_{R}}{\pi P}\partial_{\varphi}-\frac{2T_{R}}{P}\partial_{t}]~~~~~~~~~~~~~~~~~~~~~~~~~~~~~~~~~~~~~~~~~
\end{eqnarray}

\begin{eqnarray}\label{37}
\overline{H}_{-1}=i\exp(2\pi T_{L}\varphi+2n_{L}t)[-\sqrt{\Delta}\partial_{r}~~~~~~~~~~~~~~~~~~~~~~~~~~~~~~~~~~~~~\\
\nonumber
-\frac{(n_{L}-n_{R})(r-r_{+})-(n_{L}+n_{R})(r-r_{-})}{\pi P\sqrt{\Delta}}\partial_{\varphi}~~~~~~~~~~~~~~~\\
\nonumber
+\frac{(T_{L}-T_{R})(r-r_{+})-(T_{L}+T_{R})(r-r_{-})}{ P\sqrt{\Delta}}\partial_{t}]~~~~~~~~~~~~~~
\end{eqnarray}
where $P=4(T_{L}n_{R}-T_{R}n_{L})$.

The quadratic Casimir is

\begin{eqnarray}\label{38}
H^2=\widetilde{H}^2=-H_{0}^2+\frac{1}{2}(H_1H_{-1}+H_{-1}H_{1})=\frac{1}{4}(y^2\partial_{y}^2-y\partial_{y})+y^2\partial_{+}\partial_{-}
\end{eqnarray}

The crucial observation is that these Casimir, when written in term of $ \varphi$ ,t and r reduces to the radial equation (\ref{11})

\begin{eqnarray}\label{39}
\widetilde{H}^2R(r)=H^2R(r)=l(l+1)R(r)
\end{eqnarray}

\begin{eqnarray}\label{40}
   [\partial_{u}(u-u_+)(u-u_{*}) \partial_{u})-\frac{u_{+}-u_{-}}{4(u-u_{+})}(\frac{\partial_{t}}{Kk_{+}}
   +\frac{2\Xi_a\Omega_{a}}{Kk_{+}}\partial_{\varphi})^2\\
     \nonumber
   +\frac{u_{+}-u_{-}}{4(u-u_{-})}(\frac{\partial_{t}}{Kk_{-}}
   +\frac{2\Xi_a\Omega_{b}}{Kk_{+}}\partial_{\varphi})^2]\Phi=l(l+1)\Phi
\end{eqnarray}\\

\section{Entropy from the CFT  }
The microscopic entropy of the dual CFT can be computed by the Cardy formula which matches with the black hole Bekenstein-Hawking entropy

\begin{eqnarray}\label{41}
S_{CFT}=\frac{\pi^2}3({C_{L}T_{L}+C_{R}T_{R}})~
\end{eqnarray}

We will assume that away from extremality, the CFT along the $\varphi$-direction will have both a left and right-moving piece, both with identical central charge. The central charge of charged of black hole (\ref{1}) was computed in  \cite{11}. In the extremal limit we have

\begin{eqnarray}\label{42}
C_{L}=C_{R}=\frac{3\pi}{2}b\frac{(r_{+}^2+a^2)(r_{+}^2+b^2)+abq}{\Xi_bKr_{+}^2}
\end{eqnarray}

From the central charges(\ref{42}) and temperatures (\ref{21}) and using equation (\ref{41}) we have

\begin{eqnarray}\label{43}
S=\frac{[\pi^2(r_{+}^2+a^2)(r_{+}^2+b^2)+abq]}{2\Xi_a\Xi_br_{+}}
\end{eqnarray}
which agrees precisely with the gravity result presented in \cite{7}.\\
\section{Hidden conformal symmetry of black hole in five dimensional ungauged supergravity }
In this section we consider hidden conformal symmetry of black hole in five dimensional ungauged supergravity , so we consider metric (\ref{1}) with $g=0$. The non-vanishing metric components are given by
\begin{eqnarray}\label{44}
   g_{00}=-1+\frac{2m\rho^2-q^2}{\rho^4}~~~~~~~~~~~~~~~~~~~~~~~~~~~~~~~~~~~~~~~~~~~~~~\\
\nonumber
   g_{03}=-\frac{a(2m\rho^2-q^2)+bq\rho^2\sin^2(\theta)}{\rho^4}~~~~~~~~~~~~~~~~~~~~~~~~~~~~~\\
\nonumber
   g_{04}=-\frac{b(2m\rho^2-q^2)+aq\rho^2\sin^2(\theta)}{\rho^4}~~~~~~~~~~~~~~~~~~~~~~~~~~~~~\\
\nonumber
   g_{33}=(r^2+a^2)\sin^2(\theta)+\frac{a[a(2m\rho^2-q^2)+2bq\rho^2]\sin^4(\theta)}{\rho^4}~~\\
\nonumber
   g_{44}=(r^2+b^2)\cos^2(\theta)+\frac{b[b(2m\rho^2-q^2)+2aq\rho^2]\cos^4(\theta)}{\rho^4}~~\\
\nonumber
   g_{34}=\frac{[ab(2m\rho^2-q^2)+(a^2+b^2)q\rho^2]\sin^2(\theta)\cos^2(\theta)}{\rho^4}~~~~~~~~\\
\nonumber
  g_{11}=\frac{\rho^2}{\Delta}~~~~~~~~~~~~~~g_{22}=\rho^2~~~~~~~~~~~~~~~~~~~~~~~~~~~~~~~~~~~~~~~
\end{eqnarray}

 For the above metric $\sqrt{-g}=r\rho^2\sin\theta\cos\theta$, the event horizons of the space-time are given by the singularities of the metric function
which are the real roots of $r^2\Delta=0$. In the low frequency limit and in near horizon region the radial equation for black holes in minimal five-dimensional ungauged  supergravity should be simplified ( $m_{\psi}=0$)

\begin{eqnarray}\label{45}
   [\partial_{u}(\Delta \partial_{u})+\frac{u_{+}-u_{-}}{4(u-u_{+})}(\frac{\omega}{k_{+}}-\frac{2\Omega_{a}}{k_{+}}m_{\varphi})^2\\
   \nonumber
   -\frac{u_{+}-u_{-}}{4(u-u_{-})}(\frac{\omega}{k_{-}}
    -\frac{2\Omega_{b}}{k_{+}}m_{\varphi})^2]\Phi=l(l+1)\Phi
\end{eqnarray}

It is easy to see the radial part of equation of motion  could be rewritten as $SL(2,R)$ with the identification  Casimir (\ref{38})
\begin{eqnarray}\label{46}
T_L=\frac{(r_++r_{-})}{ 2\pi b}~~~~~~~~~~~~~T_{R}=\frac{(r_+-r_{-})}{ 2\pi b}
  \end{eqnarray}

\begin{eqnarray}\label{47}
n_L=-\frac{r_{+}^2}{2b}(a+b)(r_{+}+r_{-})[(r_+^2+a^2)(r_+^2+b^2)+abq]
  \end{eqnarray}

\begin{eqnarray}\label{48}
n_L=-\frac{ r_{+}^2}{2b}(a-b)(r_{+}-r_{-})[(r_+^2+a^2)(r_+^2+b^2)+abq]
  \end{eqnarray}
These results can be obtained from Eqs.(\ref{21})-(\ref{23}) by substituting $g=0$.
The microscopic entropy of the dual CFT can be computed by the Cardy formula which matches with the black hole Bekenstein-Hawking entropy.  The central charge of charged black hole in minimal five dimensional supergravity  was computed in  \cite{7}. In the extremal limit we have

\begin{eqnarray}\label{49}
C_{L}=C_{R}=\frac{3\pi}{2}b\frac{(r_{+}^2+a^2)(r_{+}^2+b^2)+abq}{r_{+}^2}
\end{eqnarray}

From the central charges (\ref{49}) and temperatures (\ref{46}) and using equation (\ref{41}) we have

\begin{eqnarray}\label{50}
S=\frac{[\pi^2(r_{+}^2+a^2)(r_{+}^2+b^2)+abq]}{2r_{+}}
\end{eqnarray}
which agrees precisely with the gravity result presented in \cite{7}.

\section{Absorption cross section}

Now we show that the absorption cross-section for scalars (with $m_{\psi}=0$) in the near-region of the black hole in minimal five dimensional supergravity  be interpreted as arising from a CFT at the left and right temperatures.
The absorption cross section is easy enough to write down using results in \cite{8}

\begin{eqnarray}\label{51}
P_{abc}\sim\sinh(\frac{\beta_{H}}{2}(\omega -m_{\varphi}\Omega_{a})\mid\Gamma(l_0+1-i\frac{\beta_{L}\omega-\beta_{H}m_{\varphi}(\Omega_{a}-\Omega_{b})}{4\pi})\mid^2\\
\nonumber
\times\mid\Gamma(l_0+1-i\frac{\beta_{L}\omega-\beta_{H}m_{\varphi}(\Omega_{a}+\Omega_{b})}{4\pi})\mid^2~~~~~~~~~~~~~~~~~~~~~~~~~
\end{eqnarray}
Where

\begin{eqnarray}\label{52}
\beta_{H}=\frac{2\pi}{k_{+}}~~~~~\beta_{R}=\frac{2\pi}{k_{+}}+\frac{2\pi}{k_{-}}~~~~~\beta_{L}=\frac{2\pi}{k_{+}}-\frac{2\pi}{k_{-}}
~~~~~l_0=\frac{l}{2}
\end{eqnarray}

To see explicitly that $P_{abs}$ matches with the microscopic greybody factor of the dual CFT, we need to identify the related parameters. From the first law of black hole thermodynamics

\begin{eqnarray}\label{53}
T_{H}\delta S=\delta M -\Omega_{a}\delta J_{\varphi}-\Omega_{b}\delta J_{\psi}+\phi\delta q
\end{eqnarray}

One can compute the conjugate charges as

\begin{eqnarray}\label{54}
\delta S_{BH}=\delta S_{CFT}=\frac{\delta E_{L}}{T_{L}}+\frac{\delta E_{R}}{T_{R}}
\end{eqnarray}
So we can get

\begin{eqnarray}\label{55}
\delta\omega_{L}=\delta E_{L}=-\frac{(a-b)m_{\phi}}{2b}+\frac{(ab+q+r_{+}^2)[(a^2+r_{+}^2)(b^2+r_{+}^2)+abq]\omega}{2br_{+}^2(r_{+}^2-ab-q)}
\end{eqnarray}

\begin{eqnarray}\label{56}
\delta\omega_{R}=\delta E_{R}=-\frac{(a+b)m_{\phi}}{2b}+\frac{(r_{+}^2-ab-q)[(a^2+r_{+}^2)(b^2+r_{+}^2)+abq]\omega}{2br_{+}^2(r_{+}^2+ab+q)}
\end{eqnarray}

Now using conformal weights $(h_{L},h_{R})=(l_{0},l_0)$ and substituting equations  (\ref{55}) and (\ref{56}) into (\ref{51}), one can find
\begin{eqnarray}\label{57}
P_{abs}\sim T_{L}^{2h_{L}-1}T_{R}^{2h_{R}-1}\sinh(\frac{\omega_{L}}{2T_{L}}+\frac{\omega_{R}}{2T_{R}})\mid\Gamma(h_{L}+i\frac{\omega_{L}}{2\pi T_{L}})\mid^2\\
\nonumber
 \times\mid\Gamma(h_{R}+i\frac{\omega_{R}}{2\pi T_{R}})\mid^2~~~~~~~~~~~~~~~~~~~~~~~~~~~~~~~~~~~~~~~~~~~~~~~
\end{eqnarray}\\
\section{Conclusion}
In this paper, we have investigate the hidden 2D conformal symmetries in the non-extremal charged rotating black holes in five dimensional gauged (ungauged) supergravity, by analyzing the near region wave equation of scalar field at low frequencies. Our result show that the absorption cross section is in consistence with the finite temperature absorption cross section for a 2D CFT. The entropy of black hole is reproduce by the Cardy formula. The authors of  \cite{10}, have shown there are two different individual 2D CFTs holographically dual to the Kerr-Newman black hole, coming from the corresponding two possible limits, the Kerr/CFT and Reissener-Nordstrom/CFT correspondences. The black hole which we have considered in this paper is a charge rotating black hole  with a couple angular momentum, so according to the suggestion of \cite{10}, we should have a threefold hidden conformal symmetry.  So, in our interesting background a probe scalar field at low frequencies can exhibit three different 2D conformal symmetries in its equation of motion. there are both left and right 2D CFTs for each of the circles $\phi$ and $\psi$. In both these cases, one can write the radial part of wave equation the casimir of an $SL(2,R)_{R}\times SL(2,R)_{L}$ algebra. However in the present paper we have considered only one of $J$-pictures,  i.e  $J_{\phi}$-picture, so we have taken the case with $m_{\psi}=0$. Moreover if one consider the charged scalar field  with  $m_{\phi}=m_{\psi}=0$, the radial part of the wave equation can be written again as the casimir of  $SL(2,R)_{R}\times SL(2,R)_{L}$  \cite{10}.


\begin{thebibliography}{99}


\bibitem{1} M. Guica, T. Hartman, W. Song and A. Strominger, "The Kerr/CFT Correspondence", arXiv: 0809.4266[hep-th].

\bibitem{2} I. Bredberg, T. Hartman, W. Song and A. Strominger, "Black Hole Superradiance
From Kerr/CFT", arXiv:0907.3477[hep-th].


\bibitem{3}O. J. C. Dias, H. S. Reall and J. E. Santos, JHEP 0908 (2009) 101;
A. J. Amsel, G. T. Horowitz, D. Marolf and M. M. Roberts, JHEP 0909 (2009) 044;
Y. Matsuo, T. Tsukioka and C. M. Yoo, Nucl. Phys. B 825 (2010) 231; "Yet Another
Realization of Kerr/CFT Correspondence", [arXiv:0907.4272 [hep-th]];
A. J. Amsel, D. Marolf and M. M. Roberts, "On the Stress Tensor of Kerr/CFT",
[arXiv:0907.3477[hep-th]];
A. Castro and F. Larsen, JHEP 0912 (2009) 037.

\bibitem{4}J. D. Brown, and M. Henneaux, Commun. Math. Phys.104,207,(1986).

\bibitem{5}T. Hartman and A. Strominger. JHEP, 0904,026,(2009)

\bibitem{6}A. Castro, A. Maloney and A. Strominger, "Hidden Conformal Symmetry of the
Kerr Black Hole", arXiv: 1004.0996[hep-th].

\bibitem{7}Z.-W. Chong, M. Cveti¡c, H. L¨u, C.N. Pope1, "General Non-Extremal Rotating Black Holes in Minimal Five-Dimensional Gauged
Supergravity", arXiv:0506029v1[hep-th].
\bibitem{Chen}
  B.~Chen and J.~Long,
  ``On Holographic description of the Kerr-Newman-AdS-dS black holes,''JHEP {\bf 1008}, 065 (2010),   arXiv:1006.0157 [hep-th].

\bibitem{8}M. Cvetic and F. Larsen, "General rotating black holes in string theory: Greybody factors
and event horizons," Phys. Rev. D 56, 4994 (1997) [arXiv:hep-th/9705192].

\bibitem{9}C. Krishnan, "Hidden Conformal Symmetries of Five-Dimensional Black Holes",
[arXiv: 1004.3537[hep-th]]; C. M. Chen and J. R. Sun, "Hidden Conformal Symmetry of the Reissner-Nordstrom
Black Holes", [arXiv: 1004.3963[hep-th]]; Y. Q. Wang and Y. X. Liu, "Hidden Conformal Symmetry of the Kerr-Newman Black
Hole", [arXiv: 1004.4661[hep-th]]; B. Chen and J. Long, "Real-time Correlators and Hidden Conformal Symmetry in
Kerr/CFT Correspondence", [arXiv: 1004.5039[hep-th]];
D. Chen, P. Wang and H. Wu, "Hidden conformal symmetry of rotating charged
black holes,"[ arXiv:1005.1404 [gr-qc]];
H. Wang, D. Chen, B. Mu, and H. Wu, "Hidden conformal symmetry of the
Einstein-Maxwell-Dilaton-Axion black hole"[ arXiv:1006.0439v1 [gr-qc]]; M. Becker, S. Cremonini and W. Schulgin, "Correlation Functions and Hidden
Conformal Symmetry of Kerr Black Holes", [arXiv:1005.3571 [hep-th]]; A. M. Ghezelbash, V. Kamali, M. R. Setare, "Hidden Conformal Symmetry of Kerr-Bolt Spacetimes", arXiv:1008.2189v1 [hep-th].
\bibitem{10} C. M. Chen, Y. M. Huang, J. R. Sun, M. F. Wu, and S. J. Zou, arXiv:1006.4097[hep-th]
 \bibitem{11}   D. D. K. Chow, M. Cveti, H. L¨u  and C. N. Pope, "Extremal Black Hole/CFT Correspondence in (Gauged) Supergravities", arXiv: 0812.2918v2 [hep-th]
\end{thebibliography}
\end{document}